\documentclass[a4paper]{article}

\usepackage{INTERSPEECH2022}
\usepackage{stfloats}
\usepackage{ragged2e}
\renewcommand{\raggedright}{\leftskip=0pt \rightskip=0pt plus 0cm}

\title{Fusion of Self-supervised Learned Models for MOS Prediction}
\name{Zhengdong Yang$^{1,\dagger}$, Wangjin Zhou$^{1,\dagger}$, Chenhui Chu$^{1,\ast}$, Sheng Li$^2$, \\ Raj Dabre$^2$, Raphael Rubino$^2$, Yi Zhao$^{3,*}$
\thanks{
$^\dagger$Equal contribution.
$^\ast$ Corresponding Author. 
}}
\address{
  $^1$Graduate School of Informatics, Kyoto University, Sakyo-ku, Kyoto, Japan\\
  $^2$National Institute of Information and Communications Technology (NICT), Kyoto, Japan\\
  $^3$Kuaishou Technology, Beijing, China}
\email{\{zd-yang, chu\}@nlp.ist.i.kyoto-u.ac.jp, zhou.wangjin.54r@st.kyoto-u.ac.jp,\\
\{sheng.li, raj.dabre, raphael.rubino\}@nict.go.jp, zhaoyi07@kuaishou.com}

\begin{document}

\maketitle
\begin{abstract}
We participated in the mean opinion score (MOS) prediction challenge, 2022. This challenge aims to predict MOS scores of synthetic speech on two tracks, the main track and a more challenging sub-track: out-of-domain (OOD). To improve the accuracy of the predicted scores, we have explored several model fusion-related strategies and proposed a fused framework in which seven pretrained self-supervised learned (SSL)  models have been engaged. These pretrained SSL models are derived from three ASR frameworks, including Wav2Vec, Hubert, and WavLM.  
For the OOD track, we followed the 7 SSL models selected on the main track and adopted a semi-supervised learning method to exploit the unlabeled data.
According to the official analysis results, our system has achieved \textbf{1$^{st}$ rank} in 6 out of 16 metrics and is one of the \textbf{top 3} systems for 13 out of 16 metrics. Specifically, we have achieved the highest LCC, SRCC, and KTAU scores at the system level on main track, as well as the best performance on the LCC, SRCC, and KTAU evaluation metrics at the utterance level on OOD track.  
Compared with the basic SSL models, the prediction accuracy of the fused system has been largely improved, especially on OOD sub-track.
\end{abstract}

\section{Introduction}
Currently, the evaluation of synthesized speech quality mainly relies on subjective listening tests, which are very expensive and time-consuming since they require numerous human listeners. 
Although there are many traditional objective speech quality evaluation methods~\cite{recommendation2001perceptual, beerends2013perceptual, malfait2006p, grancharov2006low}, they either require clean reference audio or are easily affected by background noise, enhancement algorithms, and application scenarios, which causes their results to differ from those obtained via manual evaluation of synthesized speech.

To improve the accuracy of objective evaluation, recently, researchers have started to explore quality assessment methods that do not require any reference audio for the application of synthetic datasets. 
Related works often employ neural network-based frameworks and use large-scale synthetic speech for training~\cite{lo2019mosnet, choi2021neural}. 
In the MOS challenge 2022, the organizers have provided three state-of-the-art approaches as baselines~\cite{huang2021ldnet, cooper2021generalization} and a large synthetic corpus collected from past Blizzard challenges~\cite{karaiskos2008blizzard,kinga2009blizzard,King2010TheBC,King2011TheBC,King2012TheBC,King2013TheBC,King2016TheBC}. The baseline systems are basically built on utterance-level MOS scores using either original speech~\cite{huang2021ldnet}, domain, or latent features~\cite{zezario2021deep, huang2021ldnet}. 

In addition to the accuracy of model fitting, one of the major challenges is that the existing system usually performs worse on out-of-domain (OOD) data. A better solution is to introduce  self-supervised learned (SSL) models.
One of the baselines provided by the organizers has proved the generalization ability of pretrained SSL models ~\cite{cooper2021generalization}. The organizers have examined the performance of fine-tuning various Wav2Vec 2.0 \cite{baevski2020wav2vec} and HuBERT \cite{hsu2021hubert} models which are pretrained using different corpora and have surprisingly found that such methods can do moderately well in even the very challenging
case of zero-shot utterance-level prediction.
However, we have observed that for pretrained models obtained from the same framework trained with different data, their performance on different metrics is not the same. 
Therefore, we suspect that the data and different model configurations used for pre-training will affect the performance.

Model fusion enables us to utilize different pretrained models within the same framework. By splicing the outputs of multiple models and feeding them into the fusion module, we can  expect that the fused framework can benefit from different models simultaneously. Fusing diverse models should lead to a large amount of combined pretrained knowledge being covered, which in turn might help achieve better results on the OOD task.

In this paper, we introduce our proposed fusion model framework.
To achieve good evaluation results, we have experimented with a number of different strategies and finally proposed an effective model fusion approach.
Compared with the simple fusion method where SSL output features are directly fed to a linear regression layer, our proposed model fusion approach achieves better performance on both the main track and OOD track.
Our model fusion approach has a significant impact on the OOD track, where we observed that  comprehensive performance improves at the expense of system-level SRCC scores.
According to the official analysis results, our system (T11) ranks first on 6 out of all 16 evaluation metrics and is one of the top 3 ranked systems on 13 out of the 16 metrics. 
On the main track, we have better performance at the system level, while on the OOD track, we are better at the utterance level.

\section{Proposed Approach}
We describe the SSL model fusion approach (fusion model) as well as a secondary approach involving fine-tuning SSL models.

\subsection{Fusion Model}
\begin{figure}[t]
	\includegraphics[scale=0.35]{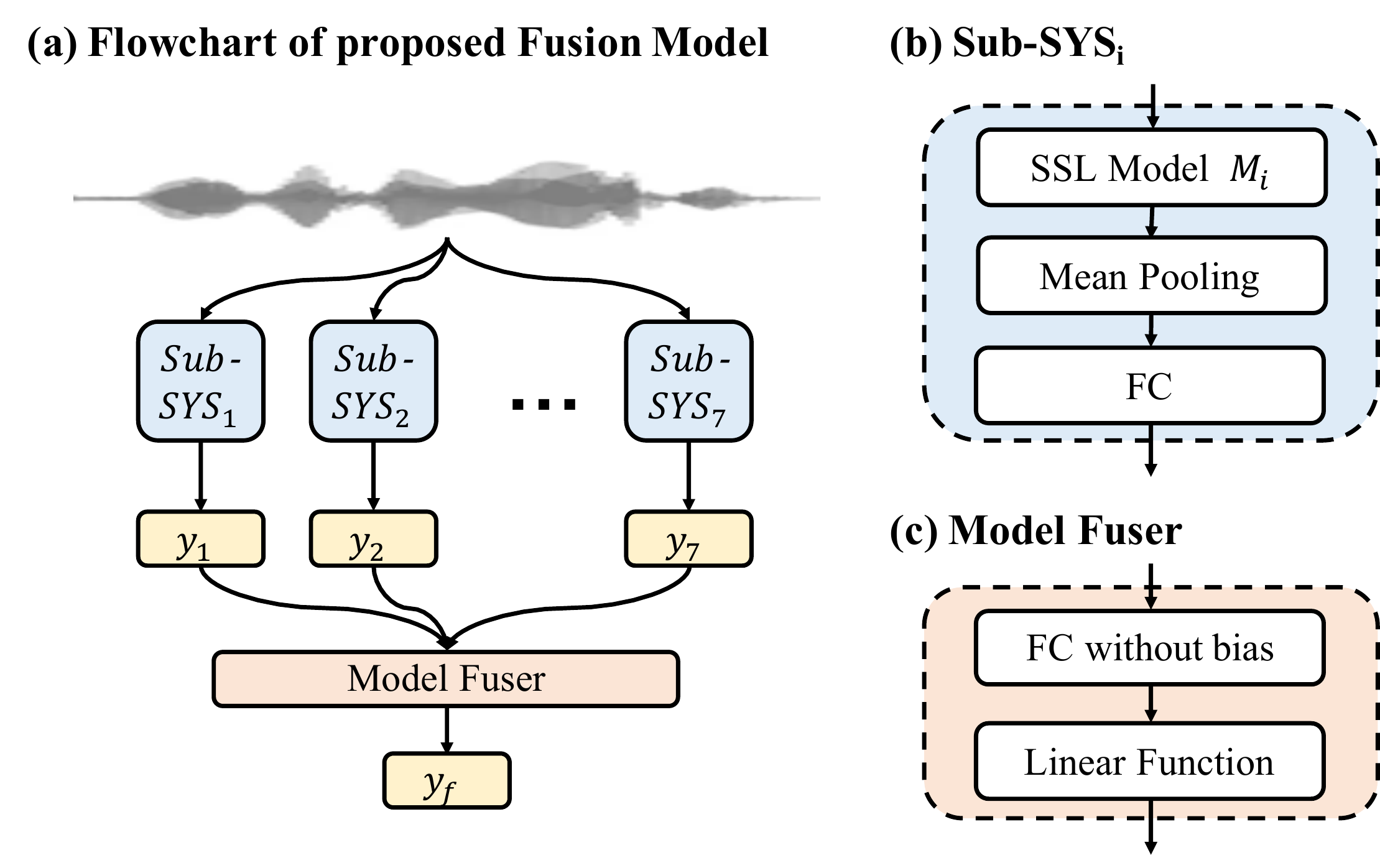}
	\caption{The flowchart of the proposed method: (a) overall structure of fusion model, (b) SSL-MOS sub-systems, (c) model fuser.}\vspace{-0.5cm}
	\label{fig:model}
	
\end{figure}

The proposed fusion model method consists of 2 separate parts: SSL-MOS sub-systems and a model fuser (as shown in Figure \ref{fig:model}). 

The challenge organizers offer a MOS scoring framework based on the self-supervised pretrained model. It fine-tunes various pretrained SSL models such as Wav2Vec 2.0 and HuBERT by mean-pooling the model’s output embeddings, adding a linear output layer, and training with the L1 loss~\cite{cooper2021generalization}. Initially, we have tried to modify the fully connected layer of this sub-system, attempting to transform the regression problem in the original model into a classification problem. However, no satisfactory results were obtained, so this paper follows the original framework as a subsystem (as shown in the Figure~\ref{fig:model} (b)). Several newly developed self-supervised pretrained models are selected and experimented as MOS-scoring subsystems. Through each subsystem, MOS scores have been predicted. Although they do not show very significant differences in evaluating metrics such as MSE at the macro level, some deviations can be found in their specific scoring. By looking at the MOS scores, it can be assumed that each sub-system captures different information from the training dataset. So the question that naturally arises is: how to combine these scores in order to get more information from the training set?

Several model fusion methods have been tested for the MOS scores combination target. By analyzing the performance of the fusion methods, a simple yet effective model fuser has been proposed. A 2-layer model has been designed (as shown in the Figure~\ref{fig:model} (c)), consisting of a fully connected layer without bias for capturing the weighted information and a linear function for obtaining the residual information between ground truth and the predicted scores after a fully connected layer.

\subsection{Fine-tuning SSL Models With ASR Evaluation}
In addition to the method above, we also seek to use the text for generating the synthesized speech as additional information to improve the accuracy of MOS prediction. The intuition is that the better the quality of the synthesized speech, the easier an ASR (Automatic Speech Recognition) system trained on human speech data can recognize the synthesized speech. However, only using ASR evaluation metrics such as Character Error Rate (CER) as inputs does not produce satisfactory results on MOS prediction because the correlation between MOS and ASR accuracy is relatively weak. We instead combine them with the method of fine-tuning SSL models.

\begin{figure}[t]
    \vspace{-0.2cm}
	\includegraphics[scale=0.5]{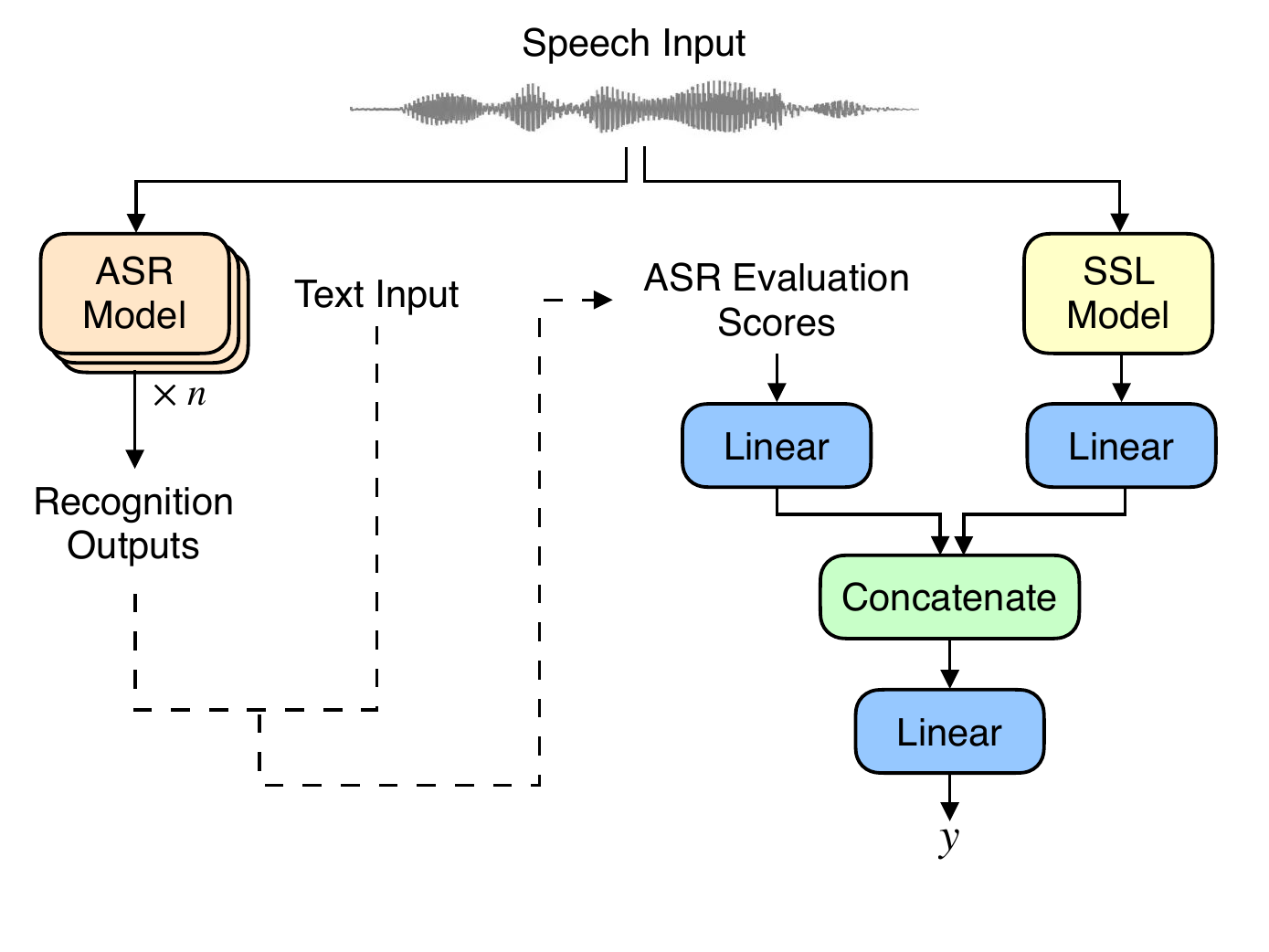}\vspace{-0.5cm}
	\caption{Framework for fine-tuning SSL models with ASR evaluation scores.}
	\label{model_cer}
\vspace{-0.6cm}
\end{figure}

The system is a modified version of the sub-system in Figure~\ref{fig:model} (b) and its structure is shown in Figure~\ref{model_cer}. We first perform ASR with different pretrained models without fine-tuning on the MOS dataset. Then the model takes both
the speech and ASR evaluation scores as inputs and takes the MOS as outputs. Similar to the fine-tuned SSL method, the model contains a pretrained SSL model to transform the speech input into contextual representation. The representation and the ASR evaluation scores then go through two separate linear layers and produce two 1-dimensional outputs. These two outputs are concatenated into a vector and go through another linear layer to predict the MOS.

\section{Datasets}

The datasets, which are divided into a main track and an “out-of-domain” (OOD) sub-track~\cite{huang2022voicemos}, are provided by the VoiceMOS competition organizers.

\subsection{Main Track}

The samples of the main track dataset come from a collection of Blizzard Challenges (BC) ~\cite{karaiskos2008blizzard,kinga2009blizzard,King2010TheBC,King2011TheBC,King2012TheBC,King2013TheBC,King2016TheBC} and Voice Conversion Challenges (VCC)~\cite{toda2016voice,wester2016analysis,lorenzo2018voice,yi2020voice,zhao2020voice}, as well as published samples from ESPnet-TTS~\cite{watanabe2018espnet}. The main track data is in English and consists of 4,974 examples for training, 1,066 examples for validation, and 1,066 examples for testing (the label of the test dataset was not revealed until the end of the VoiceMOS Challenge). There are 33 types of MOS scores on the main track, with scores ranging from 1 to 5 in steps of 0.125. 

\subsection{Out-of-domain Track}

\begin{figure}[t]
    \vspace{-0.2cm}
	\includegraphics[scale=0.4]{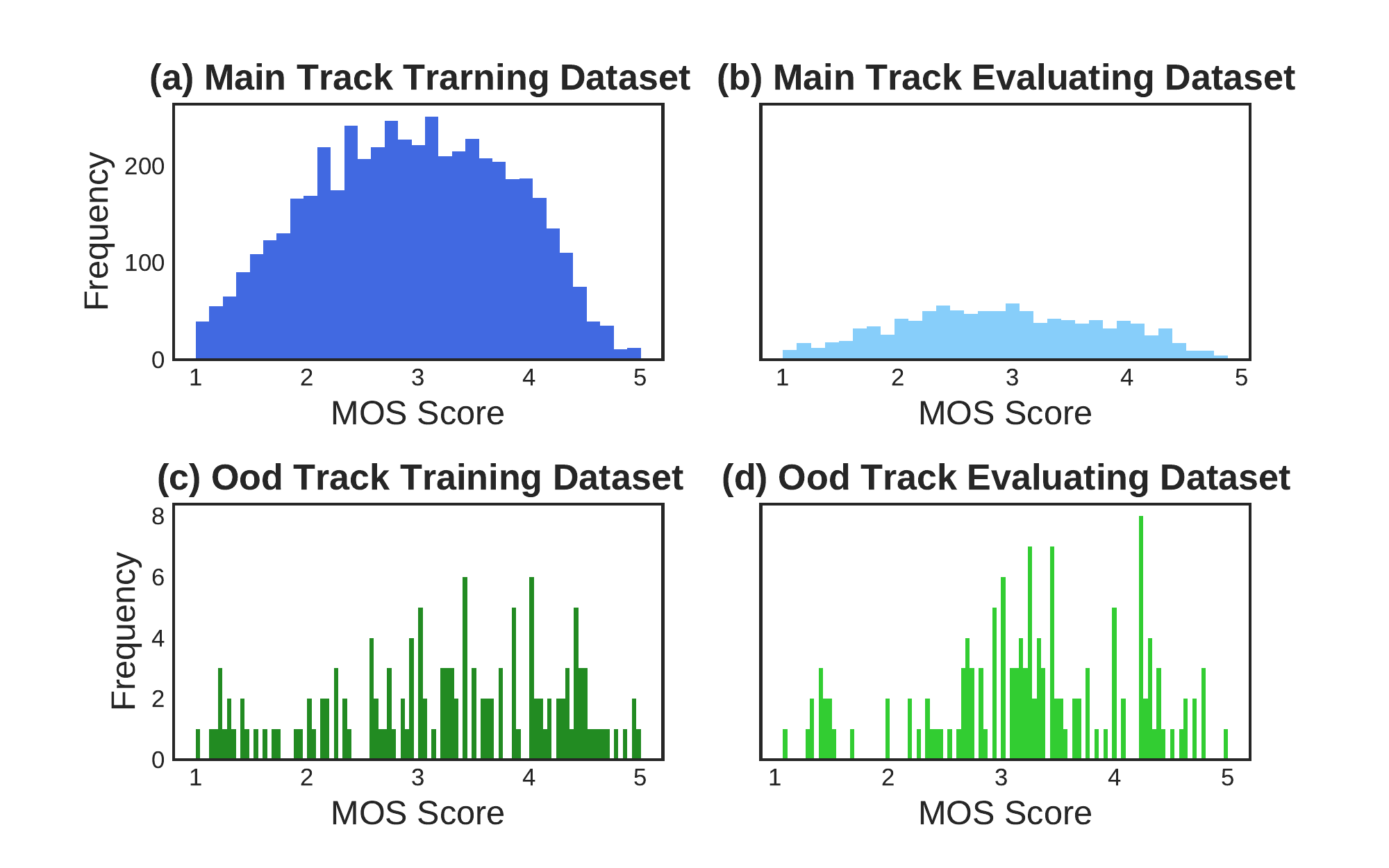}
	\caption{The MOS score distributions: (a) training dataset of the main track, (b) evaluation dataset of the main track, (c) training dataset of the OOD track, (d) evaluation dataset of the OOD track.}
	\label{fig:data}
	\vspace{-0.6cm}
\end{figure}

The out-of-domain (OOD) track is  
intended to test the generalization ability of the model. The samples of the OOD track come from the listening test of the Blizzard Challenge 2019~\cite{wu2019blizzard} (the audio samples from this challenge were not included in our main track listening test). Different from the main track, the OOD track data is in Chinese. It has provided 136 labeled examples for training, 136 examples for validation, and 540 examples for testing (the label of the test dataset was not revealed until the end of the VoiceMOS Challenge). Moreover, 540 unlabeled examples are provided for training  as well. Note that the MOS scores of the labeled samples in the OOD track cannot be categorized into 33 types like the main track. This is one of the reasons why we don't use the classification method as our sub-system. Figure~\ref{fig:data} shows the MOS score distributions on both tracks.

\section{Experiments}
\label{sec:exp-con}

\begin{table*}[!t]
\footnotesize
\centering
\caption{Results on the main track. Models finally
     selected for fusion are marked in bold.}\vspace{-0.3cm}
    \label{tab_ssl}
\begin{tabular}{l|rrrr|rrrr}
\hline \multicolumn{9}{c}{\small (a) Results of fine-tuning different pretrained SSL models individually for MOS prediction.}\\
\hline & \multicolumn{4}{c}{ Utterance level } & \multicolumn{4}{|c}{ System level } \\
Pretrained SSL Model & MSE & LCC  & SRCC & KTAU & MSE & LCC & SRCC & KTAU \\
\hline \textbf{W2V 2.0 Base} & $0.235$ & $0.875$ & $0.878$ & $0.707$ & $0.094$ & $0.935$ & $0.941$ & $0.803$ \\
\textbf{W2V 2.0 Large} & $0.197$ & $0.875$ & $0.873$ & $0.697$ & $0.068$ & $0.948$ & $0.953$ & $0.820$ \\
\textbf{W2V 2.0 Large (LV-60)} & \boldmath{$0.191$} & $0.878$ & $0.878$ & $0.704$ & \boldmath{$0.060$} & \boldmath{$0.950$} & $0.953$ & $0.823$ \\
\textbf{HuBERT Base} & $0.207$ & $0.878$ & $0.876$ & $0.700$ & $0.077$ & $0.944$ & $0.947$ & $0.812$ \\
HuBERT Large & $0.288$ & $0.813$ & $0.809$ & $0.623$ & $0.103$ & $0.923$ & $0.924$ & $0.757$ \\
HuBERT Extra Large & $0.229$ & $0.852$ & $0.849$ & $0.666$ & $0.082$ & $0.930$ & $0.931$ & $0.777$\\
\textbf{WavLM Base} & $0.199$ & \boldmath{$0.891$} & \boldmath{$0.891$} & \boldmath{$0.722$} & $0.072$ & $0.949$ & $0.954$ & $0.828$ \\
\textbf{WavLM Base+} & $0.248$ & $0.879$ & $0.883$ & $0.709$ & $0.115$ & $0.948$ & \boldmath{$0.958$} & \boldmath{$0.832$} \\
\textbf{WavLM Large} & $0.192$ & $0.876$ & $0.872$ & $0.695$ & $0.063$ & \boldmath{$0.950$} & $0.952$ & $0.827$ \\
Data2Vec & $0.314$ & $0.826$ & $0.842$ & $0.660$ & $0.144$ & $0.905$ & $0.931$ & $0.779$ \\
\hline \multicolumn{9}{c}{\small (b) Results on the main track validation set with different fusion methods. }\\
\hline & \multicolumn{4}{c}{ Utterance level } & \multicolumn{4}{|c}{ System level } \\
Fusion Method & MSE & LCC  & SRCC & KTAU & MSE & LCC & SRCC & KTAU \\
\hline LightGBM & $0.187$ & $0.885$ & $0.883$ & $0.709$ & $0.055$ & $0.953$ & $0.956$ & $0.827$ \\
Neural Networks & $0.172$ & $0.890$ & $0.891$ & $0.719$ & $0.055$ & $0.956$ & $0.960$ & $0.840$ \\
Voting & $0.167$ & \boldmath{$0.902$} & \boldmath{$0.902$} & \boldmath{$0.736$} & $0.054$ & $0.957$ & $0.961$ & $0.848$ \\
Weighted Voting & $0.164$ & $0.898$ & $0.897$ & $0.728$ & \boldmath{$0.049$} & $0.958$ & $0.961$ & $0.845$ \\
Linear Regression & $0.160$ & $0.898$ & $0.898$ & $0.730$ & $0.052$ & \boldmath{$0.961$} & \boldmath{$0.965$} & \boldmath{$0.853$} \\
\textbf{Linear Regression (features)} & \boldmath{$0.154$} & \boldmath{$0.902$} & $0.901$ & $0.735$ & $0.057$ & $0.956$ & $0.957$ & $0.841$ \\
\textbf{Proposed Model Fuser} & $0.156$ & \boldmath{$0.902$} & $0.901$ & $0.735$ & $0.051$ & $0.960$ & $0.962$ & $0.848$ \\
Linear Regression (with CER) & $0.160$ & $0.899$ & $0.898$ & $0.730$ & $0.053$ & $0.960$ & $0.964$ & $0.851$\\
Proposed Model Fuser (with CER) & $0.157$ & $0.901$ & $0.901$ & $0.734$ & $0.055$ & $0.958$ & $0.963$ & $0.850$ \\
\hline
\end{tabular}
\vspace{-0.3cm}
\end{table*}

\subsection{Fusion Model: Performance On Main Track}
\label{subsec:mod_fu}

For the main track, we first examined 10 pretrained SSL models~\cite{baevski2020wav2vec,chen2021wavlm,hsu2021hubert,chen2021wavlm,baevski2022data2vec} for MOS prediction individually, and the results are shown in Table~\ref{tab_ssl} (a). 
And then 7 well-performed SSL models (Wav2Vec 2.0 Base, Wav2Vec 2.0 Large, Wav2Vec 2.0 (LV-60), HuBERT Base, WavLM Base, WavLM Base+ and WavLM Large) are chosen for model fusion. 
In the preparation stage, 
each sub-system is firstly fine-tuned independently on the same training dataset. During the fine-tuning period, the L1 loss is used, and the model is trained for 1,000 epochs with a batch size of 1 and a learning rate of 0.0001, and early stopping is adopted if the loss hasn’t decreased for 20 epochs. Then, the model with the lowest validation loss is chosen.
Next, five basic fusion methods (voting, weighted voting, neural networks, linear regression, and random forests (LightGBM~\cite{ke2017lightgbm} in particular)) have been tested as the model fuser (as shown in the Table~\ref{tab_ssl} (b)). The weights in weighted voting are learned during training. And the neural networks use Sigmoid as activation function.
From the results, we found voting is better than LightGBM and neural networks with the sigmoid activation function, which indicates that the non-linear information is less significant than linear information. 
We also found weighted voting performs better than voting, from which we can infer that the contribution of different sub-models to the overall score is different. In addition, linear regression is better than weighted voting, we suppose that linear regression might be able to retain the residual information of the ground truth. 
Our model fuser is implemented based on the above descriptions. 

\begin{figure}[!t]
	\includegraphics[scale=0.3]{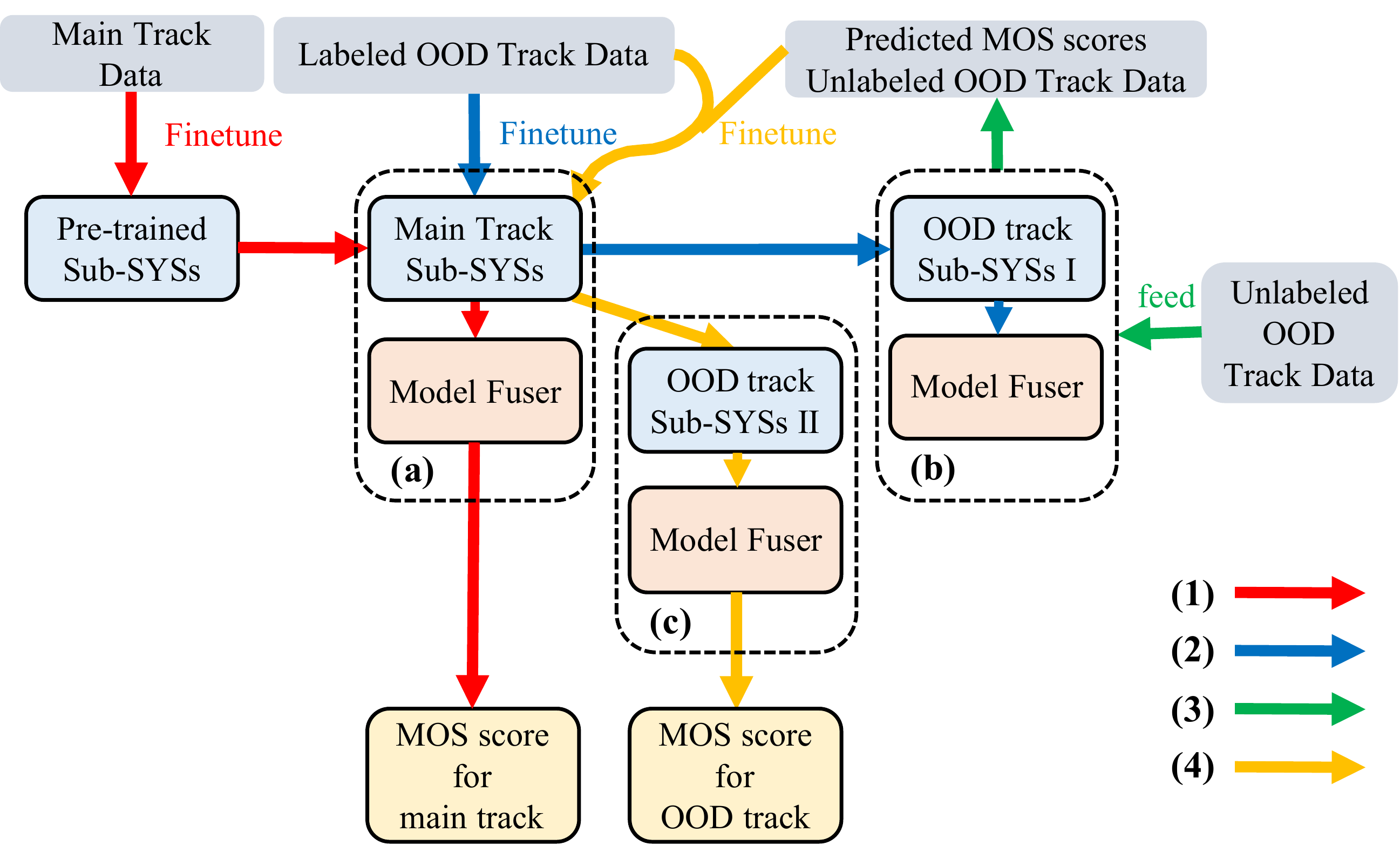}
	\caption{Implementation details of our proposed approach: (a) main track fusion model, (b) OOD track fusion model \uppercase\expandafter{\romannumeral1}, (c) OOD track fusion model \uppercase\expandafter{\romannumeral2}}
	\label{implementation_details}
	\vspace{-0.5cm}
\end{figure}

The above model fuser is an attempt to use the MOS scores predicted by each sub-model. However, each sub-system would compress the information of the latent features obtained from the SSL pretrained model in the process of predicting MOS scores. Therefore, we also tried to directly concatenate output features of each SSL pre-model and then run regression. 
In the challenge, we submitted the results of both versions of the model fusion strategy and ultimately chose our model fuser since it gives better overall performance.

\begin{table*}[!t]
\small
    \centering
    \caption{Results on the OOD track with different fusion methods. Submission versions are marked in bold.} \vspace{-0.3cm}
    \label{tab_fusion_ood}
\begin{tabular}{lc|rrrr|rrrr}
\hline & & \multicolumn{4}{c}{ Utterance level } & \multicolumn{4}{|c}{ System level } \\
Fusion Method & Unlabeled data & MSE & LCC  & SRCC & KTAU & MSE & LCC & SRCC & KTAU \\
\hline Voting & - & $0.169$ & $0.892$ & $0.783$ & $0.597$ & $0.050$ & $0.978$ & $0.948$ & $0.812$ \\
Linear Regression & - & $0.176$ & $0.886$ & $0.768$ & $0.584$ & $0.040$ & $0.979$ & $0.947$ & $0.819$ \\
Linear Regression (features) & - & $0.166$ & $0.893$ & $0.783$ & $0.598$ & $0.039$ & $0.981$ & $0.956$ & $0.841$ \\
Proposed Model Fuser & - & $0.171$ & $0.890$ & $0.779$ & $0.595$ & $0.039$ & $0.979$ & $0.959$ & $0.833$ \\
Voting & $\surd$ & $0.159$ & \boldmath{$0.903$} & $0.796$ & $0.609$ & $0.051$ & $0.981$ & $0.961$ & $0.845$ \\
Linear Regression & $\surd$ & $0.155$ & $0.900$ & \boldmath{$0.804$} & \boldmath{$0.619$} & $0.042$ & $0.980$ & $0.957$ & $0.833$ \\
\textbf{Linear Regression (features)} & $\surd$ & $0.158$ & $0.898$ & $0.791$ & $0.608$ & $0.040$ & \boldmath{$0.982$} & \boldmath{$0.969$} & \boldmath{$0.870$} \\
\textbf{Proposed Model Fuser} & $\surd$ & \boldmath{$0.149$} & \boldmath{$0.903$} & $0.790$ & $0.604$ & \boldmath{$0.036$} & \boldmath{$0.982$} & $0.958$ & $0.841$ \\
\hline
\end{tabular}
\end{table*}

\subsection{Semi-Supervised Approach: OOD Track Performance}

\begin{table*}[!t]
\small
    \centering
    \vspace{-0.1cm}
    \caption{Results on the validation set with different sets of CERs as inputs. To obtain CERs, eight Wav2Vec 2.0 models fine-tuned for the ASR task are used to conduct ASR, including 4 different model settings (Wav2Vec 2.0 Base, Wav2Vec 2.0 Large, Wav2Vec 2.0 Large LV-60 and Wav2Vec 2.0 Large LV-60 with self-training) and 2 different fine-tuning data settings (100 hours and 960 hours of Librispeech data). We experimented with using single or multiple CER inputs (predicted by two Wav2Vec 2.0 Base models, 4 models fine-tuned on 100 hours of Librispeech data or all of the 8 models). Wav2Vec (W2V) 2.0 base is used as the pretrained SSL model. (Note that the pretrained SSL model is different from the ASR model.)}\vspace{-0.3cm}
    \label{tab_cer}
\begin{tabular}{l|rrrr|rrrr}
\hline & \multicolumn{4}{c}{ Utterance level } & \multicolumn{4}{|c}{ System level } \\
ASR Models & MSE & LCC  & SRCC & KTAU & MSE & LCC & SRCC & KTAU \\
\hline W2V 2.0 Base (100h) & \boldmath{$0.184$} & $0.882$ & $0.883$ & $0.709$ & \boldmath{$0.067$} & \boldmath{$0.948$} & \boldmath{$0.952$} & \boldmath{$0.823$} \\
W2V 2.0 Base (960h) & $0.209$ & $0.874$ & $0.872$ & $0.697$ & $0.085$ & $0.935$ & $0.938$ & $0.798$ \\
2 models (W2V 2.0 Base) & $0.209$ & \boldmath{$0.886$} & \boldmath{$0.886$} & \boldmath{$0.714$} & $0.086$ & $0.947$ & $0.948$ & $0.817$ \\
4 models (100h) & $0.255$ & $0.865$ & $0.875$ & $0.702$ & $0.114$ & $0.923$ & $0.943$ & $0.803$ \\
8 models & $0.247$ & $0.881$ & $0.880$ & $0.704$ & $0.136$ & $0.933$ & $0.933$ & $0.797$ \\
\hline
\end{tabular}
\vspace{-0.1cm}
\end{table*}

\begin{table*}[!t]
\footnotesize
    \centering
    \caption{SRCC in utterance-level (UTT) and system-level (SYS) for the combination of different pretrained SSL models and ASR models. The results that are better than the baseline (finetuned SSL models without ASR evaluation) are marked in bold.}\vspace{-0.3cm}
    \label{tab_comb}
\begin{tabular}{l|rr|rr|rr}
\hline & \multicolumn{6}{c}{ ASR Model } \\
Pretrained SSL Model & \multicolumn{2}{c}{W2V 2.0 Base (100h)} & \multicolumn{2}{|c}{W2V 2.0 Large (100h)} & \multicolumn{2}{|c}{W2V 2.0 Large (LV-60) (100h)} \\
& UTT & SYS & UTT & SYS & UTT & SYS \\
\hline W2V 2.0 Base & \boldmath{$0.883$} & \boldmath{$0.952$} & \boldmath{$0.881$} & $0.936$ & $0.871$ & $0.935$ \\
W2V 2.0 Large & $0.846$ & $0.936$ & \boldmath{$0.878$} & \boldmath{$0.955$} & \boldmath{$0.886$} & $0.935$ \\
W2V 2.0 Large (LV-60) & $0.813$ & $0.892$ & $0.876$ & $0.951$ & $0.829$ & $0.930$ \\
HuBERT Base & $0.858$ & $0.938$ & $0.872$ & \boldmath{$0.954$} & $0.872$ & $0.943$ \\
WavLM Base & $0.880$ & $0.939$ & $0.880$ & $0.944$ & $0.884$ & $0.942$ \\
WavLM Base+ & $0.866$ & $0.935$ & $0.871$ & $0.942$ & $0.876$ & $0.936$ \\
WavLM Large & $0.868$ & $0.942$ & $0.864$ & $0.949$ & $0.869$ & \boldmath{$0.954$} \\
\hline
\end{tabular}
\vspace{-0.3cm}
\end{table*}

Because the OOD track contains unlabeled data, we attempted to predict the MOS scores of the OOD data by a semi-supervised learning method. 
As shown in the Figure~\ref{implementation_details}, the method consists of 4 steps: (1) We train our proposed system with the main track data to obtain system $a$. (2) We fine-tune the sub-systems trained in step 1 (referred to as the main track sub-systems) with only labeled OOD data and conduct model fusion with the fine-tuned sub-systems to obtain system $b$. (3) We feed the unlabeled data to system $b$ and obtain MOS scores to label them. (4) We combine the newly-labeled data in step 3 with the labeled OOD data to create a new OOD training dataset, with which we fine-tune the main track sub-systems and conduct model fusion again to obtain system $c$. MOS scores predicted by system $a$ are used for the main track and MOS scores predicted by system $c$ are used for the OOD track. 

 

Table~\ref{tab_fusion_ood} shows the results of various fusion model methods with semi-supervised learning versus those with no unlabeled data at all. It can be noticed that our semi-supervised approach improves the performance significantly.

\subsection{Fine-tuning SSL Models With ASR Evaluation}

We used CER to evaluate the accuracy of the ASR output and experimented with using single or multiple CER results produced by different ASR models, as shown in Table~\ref{tab_cer}. The results indicate that using a single CER input is generally better than using multiple CER inputs. We have examined different combinations of SSL models and ASR models, as shown in Table~\ref{tab_comb}. The results show that not every combination can improve the performance compared to the baseline (fine-tuned SSL models without ASR evaluation), which might be the reason behind the inferior performance of using multiple CER inputs.  Note that
the performance tends to be better when the same architecture is used for the pretrained SSL model and the ASR model. We finally picked out two combinations with better performance (Wav2Vec 2.0 Base with Wav2Vec 2.0 Base 100h as ASR model, Wav2Vec 2.0 Large with Wav2Vec 2.0 Large 100h as ASR model) together with other five fine-tuned SSL models without ASR evaluation for model fusion. However, the results in Table~\ref{tab_ssl} (b) shows that the fusion result is not improved compared to that without ASR evaluation involved, and the reason needs to be further investigated.

\section{Conclusion}
\label{sec:con}
For the MOS 2022 challenge, we have experimented with a number of different strategies and finally proposed an effective fusion model.
For the main track, we selected 7 pretrained SSL models as sub-systems and designed a simple yet effective model fuser based on the analysis of five model fusion methods.
For the OOD track, we followed the 7 SSL models chosen on main track and adopted a semi-supervised learning method which has improved the performance significantly.
Our final submission uses model fuser to fuse 7 subsystems on main track and uses the semi-supervised learning method on OOD track. 
We also attempted to utilize text information by fine-tuning SSL models with ASR evaluation.
Our systems ranked first in 6 out of 16 evaluation metrics over all tracks. Furthermore, in 13 out of 16 metrics, our systems were one of the top 3 ranked ones.

\newpage

\bibliographystyle{IEEEtran}
\bibliography{mybib}

\end{document}